\documentclass[useAMS,usenatbib]{mn2e}

\usepackage{natbib}
\usepackage{graphicx}
\usepackage[space]{grffile}
\usepackage{latexsym}
\usepackage{amsfonts,amsmath,amssymb}
\usepackage{url}
\usepackage[utf8]{inputenc}
\usepackage{fancyref}
\usepackage{hyperref}
\hypersetup{colorlinks=false,pdfborder={0 0 0},}

\title[Tomography of X-ray Nova Muscae 1991]{Tomography of X-ray Nova Muscae 1991: Evidence for ongoing mass transfer and stream-disc overflow$^{}$\thanks{This paper includes data gathered with the 6.5 meter Magellan Telescopes located at Las Campanas Observatory, Chile.}}
\author[Peris et al.]{Charith S. Peris,$^{1,2}$\thanks{E-mail:
cperis@cfa.harvard.edu; c.peris@neu.edu} 
  Saeqa D. Vrtilek,$^{1}$
  James F. Steiner,$^{1}$ 
  Jan M. Vrtilek,$^{1}$ 
\newauthor 
  Jianfeng Wu,$^{1}$ 
  Jeffrey E. McClintock,$^{1}$
  Penelope Longa-Pe{\~n}a,$^{3}$ 
  Danny Steeghs,$^{3}$
\newauthor 
  Paul Callanan,$^{4}$
  Luis C. Ho,$^{5,6}$
  Jerome A. Orosz,$^{7}$
  Mark T. Reynolds,$^{8}$  \\
$^{1}$Harvard-Smithsonian Center for Astrophysics, 60  Garden Street, Cambridge, MA 02138, U.S.A.\\
$^{2}$Department of Physics, Northeastern University, 360 Huntington Avenue, Boston, MA 02115, U.S.A.\\
$^{3}$Department of Physics, University of Warwick, CV4 7AL, Coventry, U.K.\\
$^{4}$Department of Physics, University College Cork, Ireland.\\
$^{5}$Kavli Institute for Astronomy and Astrophysics, Peking University, Beijing 100871, China.\\
$^{6}$Department of Astronomy, School of Physics, Peking University, Beijing 100871, China.\\
$^{7}$Department of Astronomy, San Diego State University, 5500 Campanile Drive, San Diego, CA 92182-1221, U.S.A.\\
$^{8}$Department of Astronomy, University of Michigan, 1085 S University Avenue, Ann Arbor, MI, 48109, U.S.A.}

\begin{document}

\date{Accepted ?? Received ??; in original form ??}

\pagerange{\pageref{firstpage}--\pageref{lastpage}} \pubyear{2015}

\maketitle

\label{firstpage}

\begin{abstract}
We present a spectroscopic analysis of the black hole binary Nova Muscae 1991 in quiescence using data obtained in 2009 with MagE on the Magellan Clay telescope and in 2010 with IMACS on the Magellan Baade telescope at the Las Campanas Observatory. Emission from the disc is observed in H~$\alpha$, H~$\beta$ and Ca II ($\lambda$8662). A prominent hotspot is observed in the Doppler maps of all three emission lines. The existence of this spot establishes ongoing mass transfer from the donor star in 2009-2010 and, given its absence in the 1993-1995 observations, demonstrates the presence of a variable hotspot in the system. We find the radial distance to the hotspot from the black hole to be consistent with the circularization radius. Our tomograms are suggestive of stream-disc overflow in the system. We also detect possible Ca II ($\lambda$8662) absorption from the donor star.
\end{abstract}


\section{Introduction}

Nova Muscae 1991 (GU Mus; hereafter Nova Muscae) is a low mass X-ray binary discovered in 1991 as a bright transient source by instruments on board the {\it Ginga} \citep{Makino91} and {\it GRANAT} \citep{Lund91} satellites.  It is a soft X-ray transient, a class of objects which displays changing levels of soft X-rays and infrequent, yet extreme X-ray outbursts. Its mass function ($f(M) = 3.02 \pm  0.05$ M$_{\sun}$, Wu et al. in preparation; $f(M) = 3.01 \pm  0.15$ M$_{\sun}$, \citealt{Orosz96}) is consistent with a black hole as its compact object.  

The optical counterpart of Nova Muscae was discovered by \citet*{DellaValle91} who noted a large increase in brightness to V $\sim$ 13.5 in January 1992. The optical spectra taken in 1992 displayed Balmer, He and N lines superimposed on a blue-peaked continuum which established it as the optical counterpart to the X-ray nova. 

An important feature of systems such as Nova Muscae is the interaction between the in-falling stream and the accretion disc. This interaction typically occurs in a small region also known as the ``hotspot'' and is observed in these systems both as a peak in the orbital light curve at phase $\sim0.8-0.9$ (where phase 0.0 is defined as the superior conjunction of the compact object) and as a bright spot at the same phase in tomograms (e.g., \citealt{Krzeminski65, Wood86}, \citealt*{Neilsen08}, \citealt{Calvelo09, Peris12}). \citet{Lubow76} were the first to suggest that some of the gaseous matter falling from the donor star onto the disc via the stream does not stop at the hotspot but flows over and under the disc. Smooth particle hydrodynamics (SPH) simulations (see \citealt*{Armitage96, Kunze01}) have predicted that a significant fraction of the stream can be deflected from the edge of the disc and overflow to smaller radii. In some cases \citep{Cornelisse07, Pearson06} observations of neutron star binaries have provided likely confirmation in the form of a tell-tale bright feature at phase $\sim 0.5-0.75$ in emission line Doppler maps reconstructed via the Doppler tomography technique \citep{Marsh88}. We will explore a possible stream-disc overflow in the accretion disc of Nova Muscae.

Previous optical tomography of Nova Muscae was carried out using data obtained with the 3.9 m Anglo-Australian Telescope and the 3.6 m New Technology Telescope \citep{Casares97}. \citeauthor{Casares97} did not detect a hotspot in their tomogram and suggested this may be evidence for low mass transfer rates in Nova Muscae compared to other X-ray transients of comparable orbital period.  

In this paper we present optical tomography of Nova Muscae in quiescence using data from the 6.5 m Magellan Baade and Clay telescopes. In Section 2 we describe our observations and data reduction. In Section 3 we analyse the spectrum and present the tomographic results while in Section 4 we discuss and interpret our results. In Section 5 we present our conclusions.

\section{Observations}

Nova Muscae was observed with the Magellan Echelette (MagE; \citealt{Marshall08}) spectrograph on the Magellan Clay telescope in 2009. Data were obtained over the nights of April 25--26, 33 spectra from the first night and 39 from the second night. 72 spectra, each containing 15 orders, were collected with individual exposure times of 600 s.  A slit width of 0.85$\arcsec$ was used throughout.  The resultant spectral resolution achieved is approximately 1~\AA~ at a wavelength of 5000~\AA, corresponding to $\sim$60 km s$^{-1}$ velocity resolution. The data reduction steps (bias subtraction, flat-field correction, wavelength calibration, and spectral extraction) were performed using the Carnegie Observatory pipeline\footnote{\url{http://code.obs.carnegiescience.edu/mage-pipeline}}. The sky-subtraction of the 3 highest wavelength orders of the resulting spectrum could not be performed due to limitations of the pipeline.

A set of 20 Nova Muscae observations was obtained using the IMACS instrument on the Magellan Baade telescope in 2010, over the nights of February 10, 11, and 12. Spectra were collected with exposure times of 1200 s each, using a 600 lines mm$^{-1}$ grating.  Atmospheric conditions allowed slits of 0.75$\arcsec$ (Feb. 10 and 11) and 1.5$\arcsec$ (Feb. 12). Spectral resolutions of 2.6~\AA~ and 5.2~\AA~ were achieved with these respective slits at H~$\alpha$, corresponding to velocity resolutions of $\sim$118 km s$^{-1}$ and $\sim$236 km s$^{-1}$.  Spectra were reduced using standard image reduction routines in IRAF version 2.10.

From examination of the images containing the Nova Muscae spectra we noted the presence of extended emission at H~$\alpha$ distributed along the direction of the slit on both sides of the spectrum. Diffuse Galactic H~$\alpha$ emission is a possible explanation; we have tested this conjecture by deriving an approximate flux calibration for the observed diffuse emission, and noting that this compares well with the H~$\alpha$ surface brightness at the position of Nova Muscae in the maps of \citet{Finkbeiner03}. We obtained a value for the flux of the extended H~$\alpha$ emission by comparing the average number of counts observed on both sides of the slit to that of a standard star (LTT 4816, observed for 300s with the same instrument settings as Nova Muscae; we assumed for comparison with the diffuse H~$\alpha$ that the flux from the point-source standard was diffused by seeing over approximately 1 arcsec$^2$). The flux of diffuse H~$\alpha$ around Nova Muscae was found to be ${S_{\lambda}}^{H_{2}} \approx 1.1 \times 10^{-16}$ ergs cm$^{-2}$ s$^{-1}$ arcsec$^{-2}$. The maps of \citet{Finkbeiner03} show a value of $1.36 \times 10^{-16}$ ergs cm$^{-2}$ s$^{-1}$ arcsec$^{-2}$ for Galactic diffuse H~$\alpha$ emission in the same region.

For the convenience of future observers, we have updated the coordinates of Nova Muscae using a 60~s image of the field obtained using the Bessel V filter on IMACS. We fitted a world coordinate system (WCS) to the image using imwcs version 2.8.7 and the 2MASS catalogue was used as a WCS reference. Nova Muscae was found to have a right ascension of 11~h 26~m 26.6~s and a declination of -68~d 40~m 32.87~s (the error of right ascension corresponds to 0.4~s while the error of the declination is dominated by the 2MASS uncertainty\footnote{\url{http://spider.ipac.caltech.edu/staff/hlm/2mass/overv/overv.html}} of 0.07-0.08~arcsec). A finding chart with the source marked has been included in the Appendix.

\section{Analysis}

\subsection{Spectrum of Nova Muscae}

In Figure 1, we present 10 successfully sky-subtracted orders of the mean MagE spectrum, covering a wavelength range of $\sim$3300--7300~\AA.  A low-order fit to the continuum has been subtracted to highlight line features, of which the Balmer lines are dominant.   Nova Muscae's Balmer lines display the classic double-peaked profile which is a signature of Doppler shifted emission originating from the accretion disc. The He I lines (5875 and 6678~\AA) and Fe II multiplet 42 (4924, 5018 and 5169~\AA; \citealt{Moore72}), noted previously by \citet{Orosz96}, are detected. \citeauthor{Orosz96} detected these features via isolation of the disk component of the Nova Muscae spectra by the subtraction of a K5 V template star spectrum. Since we do not isolate the disk component we find these features at lower $S/N$. Additionally, we observe the He I line at 4471~\AA. The NIII/CIII Bowen blend that \citet{Orosz96} observed after the 1992 outburst is very weak in our spectra. 

Ca II emission is detected in higher wavelength orders of the MagE spectrum which were not successfully sky-subtracted. We present one such emission line of interest ($\lambda=8662$~\AA) in Figure 2 with which we produce a tomogram. It should, however, be interpreted with caution due to the presence of unsubtracted sky-lines.

In analogy to Figure~1, Figure 3 presents the mean IMACS spectrum after continuum subtraction. Due to the low S/N we could not distinguish any features suitable for tomography apart from the strong H~$\alpha$ line. Therefore, we present only the data from IMACS/CCD2 which covered the wavelength range $\sim$6000--6700~\AA.

\begin{figure*}
\includegraphics[width=180mm]{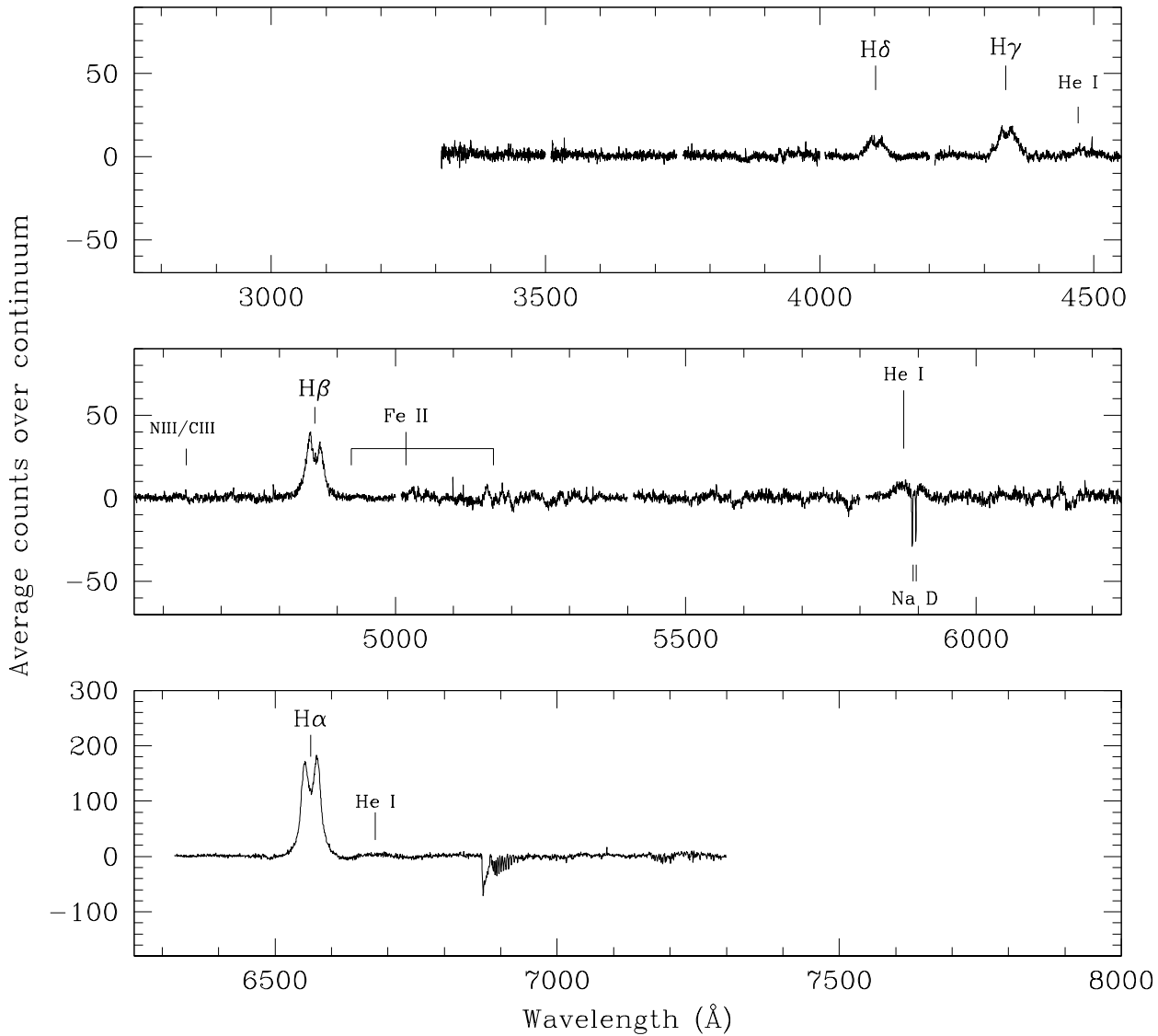}
\caption{The continuum-subtracted mean spectrum of Nova Muscae, from 10 successfully sky-subtracted orders of the MagE data with a total integration time of 720 minutes. The spectrum has not been corrected for telluric absorption. Balmer lines dominate the spectral profile.} 
\end{figure*}

\begin{figure}
\includegraphics[width=70mm]{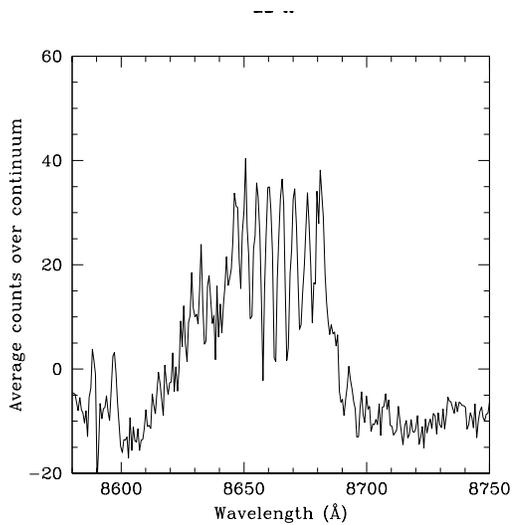}
\caption{The Ca II ($\lambda$8662) emission line obtained from an order of the MagE spectrum that was affected by sky-lines.} 
\end{figure}

\begin{figure*}
\begin{center}

\includegraphics[width=70mm]{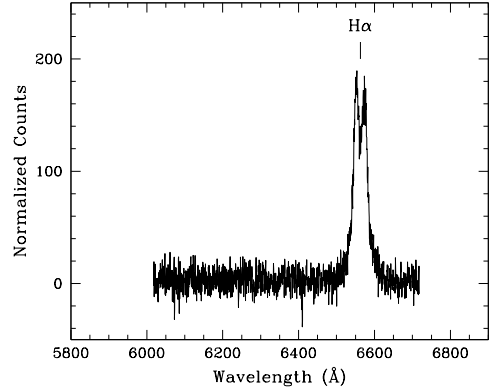}
\caption{The continuum-subtracted mean spectrum of Nova Muscae, created using IMACS/CCD2 data taken on Feb 10. The total integration time is 160 minutes.}
\end{center}
\end{figure*}

\subsection{Doppler Imaging}

Standard Doppler tomography constructs the distribution of line emission from a binary system in a two dimensional velocity space image \citep{Marsh88}. It uses the line emission at each orbital phase to calculate the line strength as a function of velocity. Figure 4 shows a schematic which makes evident the relationship between velocity coordinates and spacial coordinates for a disc-and-star system.

In order to generate the tomograms, each spectrum was binned into equal velocity bins and a linear fit performed to the continuum around each line. The continuum was then subtracted and the residual spectrum provided as input to the tomography software, into one of 60 phase interval bins.   The tomography software modifies a template starting image (a uniform grid of pixels or a two dimensional Gaussian), to fit the data by minimizing its reduced $\chi^{2}$ in comparison to the data. To break the degeneracy inherent in this approach, the maximum entropy method (MEM) is employed to select the smoothest image during computation.  

One shortcoming of Doppler tomography is that it assumes emission from all structures is continuous over the full binary orbit. Modulation Doppler tomography\footnote{For comprehensive descriptions of Doppler tomography and modulation tomography, respectively, we refer the interested reader to \citet{Marsh01} and \citet{Steeghs03}.} is an extension of standard Doppler tomography that allows the line flux from a given point to vary as a function of time enabling phase-dependent information carried by the line profiles to be explored. We produce additional modulation Doppler tomograms for each line. 

\begin{table*}
\begin{center}
\caption{System Parameters of Nova Muscae}
\label{log}
\begin{tabular}{@{}lcc}
\hline
\hline
Orbital Period, $P_{\rm orb}$ (days) & 0.43260249 $\pm$ 0.00000009\,{\it $^{a,*}$}  \\
Zero ephemeris, $T_0$ (days) & 2454946.79462 $\pm$ 0.00042\,{\it $^{b}$} \\
Systemic Velocity, $\gamma$ & 14.24 $\pm$ 2.08 km s$^{-1}$\,{\it $^{a,*}$} \\
Inclination, $i$ & 54 $\pm$ 1.5$^{\circ}$\,{\it $^{c}$}  \\
Mass Function, $f(M)$ & $3.02 \pm 0.05$ M$_{\sun}$\,{\it $^{a,*}$}  \\
Mass Ratio, $q$ & $0.068 \pm 0.003$\,{\it $^{a}$}  \\
$K_2$ & 406.83 $\pm$ 2.22 km s$^{-1}$\,{\it $^{a,*}$}  \\
\hline
\hline

{\it $^{a}$} Wu et al. (in preparation) {\it Derived using the MagE spectra used in this paper.} \\
{\it $^{b}$} {\it Derived from the spectroscopic ephemeris calculated by Wu et al.}\\
{\it $^{c}$} \citet*{Gelino01} \\
{\it $^{*}$} {\it Consistent with the values derived by \citet{Orosz96}}\\

\hline
\end{tabular}
\end{center}
\end{table*}

\begin{figure*}
\begin{center}
\includegraphics[width=160mm]{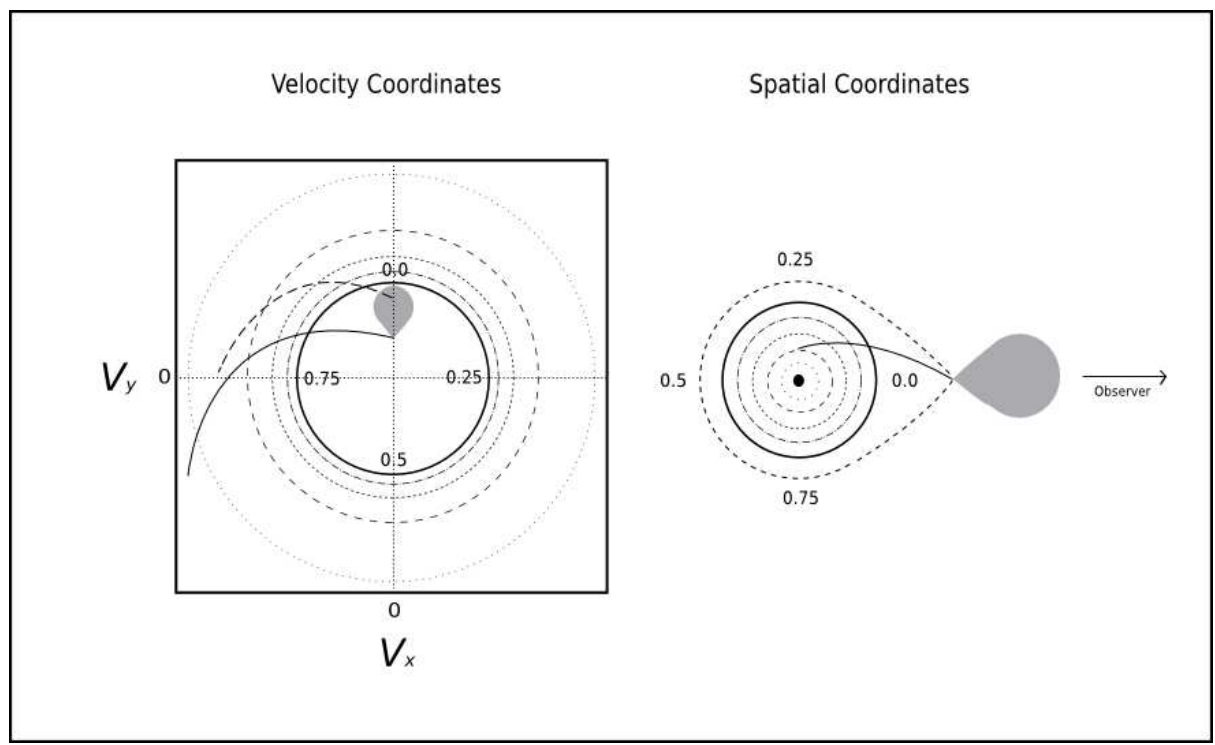}
\caption{A schematic of the relationship between velocity coordinates (left) and spatial coordinates (right) in a Doppler map. The trajectory of the accretion stream is shown as the arced solid line in both panels, and in both the Roche lobe of the secondary star is shown in solid gray.  The dashed trajectory in the left panel shows Keplerian velocities along the gas stream.   In this representation, the star orbits counter-clockwise from the perspective of a fixed observer to the right; the number at the position of the secondary star labels its phase. The solid-black circle in the right panel marks the outer edge of the disc, at which the orbital velocity in the disc is lowest. The Roche lobe of the black hole primary is represented by a dashed line. 
}\label{fig:schema}
\end{center}
\end{figure*}

\subsection{The Maps}

The orbital and binary characteristics of Nova Muscae have been studied elsewhere, with properties summarized in Table~1. Using these values as input, in Figures 5--8, we present a triplet of tomograms of Nova Muscae, each having been computed for a different spectral line: H~$\alpha$, H~$\beta$ and Ca~II.  The four panels of each figure depict the following: On the upper left, observed line profiles as a function of phase;  on the upper right, the profiles predicted from the Doppler map; on the lower left, the Doppler map;  on the lower right, the amplitude modulations that result when constant disc emission is subtracted from the Doppler map.   
 
Within the panel that contains the Doppler map, we over plot the Roche lobe of the secondary and mark two tracks representing the stream and disc velocities as in Figure~\ref{fig:schema}.  The lower track shows the ballistic trajectory of the gas stream and the upper gives the Keplerian disc velocity along the gas stream.  Each open circle marks 0.1$R_{\rm L1}$ and each point represents 0.01$R_{\rm L1}$ where $R_{\rm L1}$ is the inner Lagrangian distance \citep{marsh05}. The hotspot due to the impact of the accretion stream on the disc is usually positioned on or in-between these lines.  The positions of the Roche lobes are set using the radial velocity semi-amplitudes $K_1$ and $K_2$. Since $K_1$ has not been directly measured for Nova Muscae, we used the mass ratio ($q$) and $K_2$ (see Table 1) to obtain a value $K_1 = 27.7 \pm 1.2$~km s$^{-1}$. The main purpose for overlaying these trajectories is to interpret the position of the hotspot and its relationship to the secondary star's Roche lobe. We therefore tested this placement by shifting $K_1$ within its errors and found the displacement to be negligible.

\begin{figure}
\includegraphics[width=1.0\columnwidth, angle=-90]{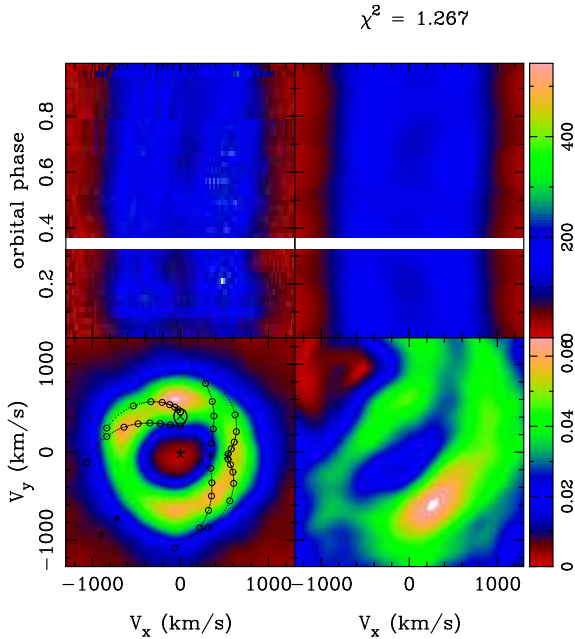}
\caption{The H~$\alpha$ maps for the MagE data in 2009. The trailed spectra is shown in the top-left panel; the Doppler map is on the bottom-left; the modulation-amplitude map is on the bottom-right; the predicted trail from the Doppler map is on the top-right.}
\label{fig:dopp1}

\end{figure}

In Figure~\ref{fig:dopp1}, we show the tomographic maps for H~$\alpha$, the strongest emission line. The disc can be seen clearly in the Doppler map on the lower left panel. Contrary to the observation by \citet{Casares97} we observe a prominent hotspot on the ballistic trajectory of the gas stream at phase $\sim$0.8 where the accretion stream impacts the disc. A crescent-shaped emission region from phases $\sim 0.2-0.6$ and an additional bright spot at phase $\sim$0.0 are also evident. These features are discussed in Section~\ref{section:disc}. 


The H~$\alpha$ emission line in the IMACS spectrum (Figure 6), which yields a similar tomogram, shows a clear hotspot as in the MagE H~$\alpha$ map (Figure 5). The bright spot at phase $\sim$0.0 is also observed. A region of excess emission corresponding to the crescent-shaped emission region manifests at phases $\sim$0.25-0.6.


The Doppler map of H~$\beta$ on the lower left hand panel of Figure 7 also displays the annular emission expected from the disc.  A bright hotspot is evident from the data at phase $\sim$0.9. The location of the hotspot in H~$\beta$ closer to the upper track, shows its velocity to be closer to that of the Keplerian disc along the gas stream than the ballistic trajectory. This is expected in a situation where the collision of the stream with the disc creates a shock which shifts the velocity of the inflowing stream toward the velocity of the disc. Such intermediate positioning has been observed in other systems (e.g., \citealt{Neilsen08}, \citealt*{Marsh94}, \citealt{Calvelo09}). 

The circular disc structure in the H~$\beta$ map extends out to higher velocities (inwards to smaller radii) than in H~$\alpha$, indicative of H~$\beta$ production by material at higher temperatures and correspondingly closer to the black hole.  The Doppler map shows no other features apart from the hotspot and the disc emission. Fluctuation of the hotspot emission harmonic to the orbital period is observed in the modulation amplitude map.

Lastly, we present the maps of Ca~II ($\lambda$8662) in Figure 8 derived from the emission line shown in Figure 2. As mentioned in Section 3.1 and evidenced by the affected intensity scale, this image should be interpreted with caution due to unsubtracted sky lines. However, the presence of clean S-curves generated by the bright spots in the predicted trail demonstrates that the features in our Doppler map are relatively unaffected by the stationary sky lines. Ca~II seems to originate from disc material at lower velocities (larger radii) when compared to the H~$\alpha$ map. This indicates that the Ca~II emission occurs closer to the outer edge of the disc, where one expects lower gas temperatures. The Doppler map in Ca~II shows the hotspot at phase $\sim$0.85. As in the case of H~$\beta$, the hotspot in Ca~II is offset from the ballistic trajectory of the gas stream. We note the presence of bright emission regions near phases $\sim$0.25 and $\sim$0.6 which correspond to the edges of the crescent-shaped emission region observed in H~$\alpha$. The hotspot and the other bright regions in Ca~II lag in phase by $\sim$0.05 when compared to the spots in the H~$\alpha$ map. We also note the presence of a compact region coincident with position of the secondary, which displays substantially lower emission relative to the surrounding region. Fluctuation of the emission from a similar location is observed in the modulation amplitude map.

\begin{figure}
\begin{center}
\includegraphics[width=1.0\columnwidth, angle=-90]{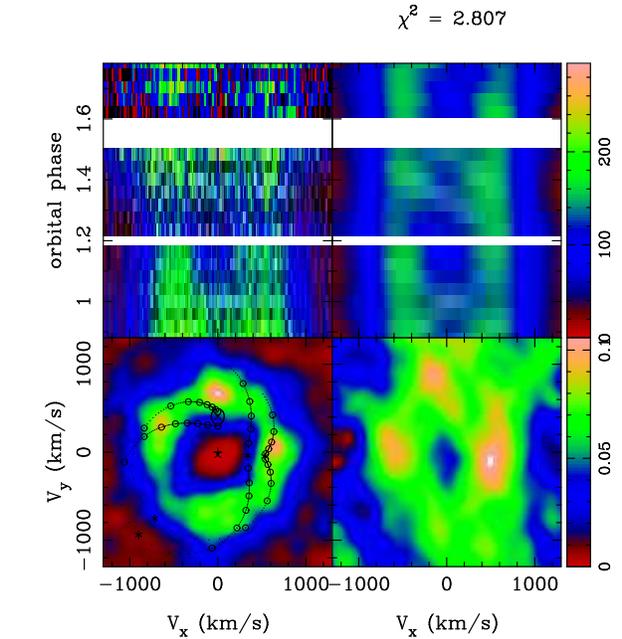}
\caption{H~$\alpha$ maps created using IMACS data taken in 2010. Panels are the same as described in Figure 5.}
\label{fig:dopp2}
\end{center}
\end{figure}

\begin{figure}
\includegraphics[width=1.0\columnwidth, angle=-90]{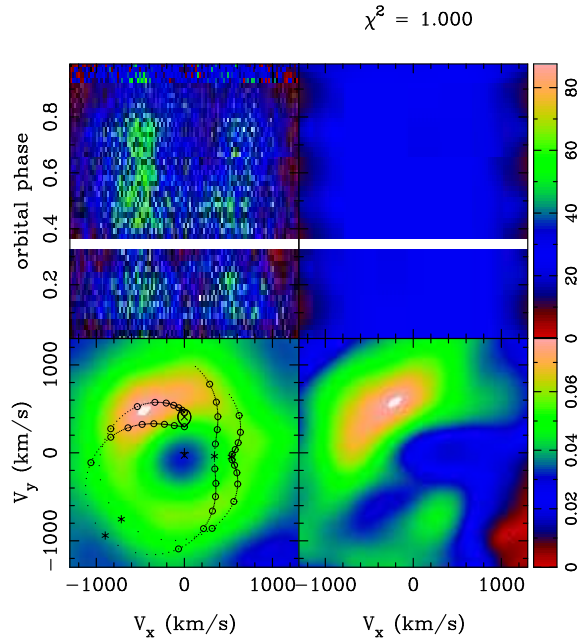}
\caption{The same as Fig.~\ref{fig:dopp1}, but for H~$\beta$, again for the MagE 2009 spectral data.}\label{fig:dopp3}
\end{figure}

\begin{figure}
\includegraphics[width=1.0\columnwidth, angle=-90]{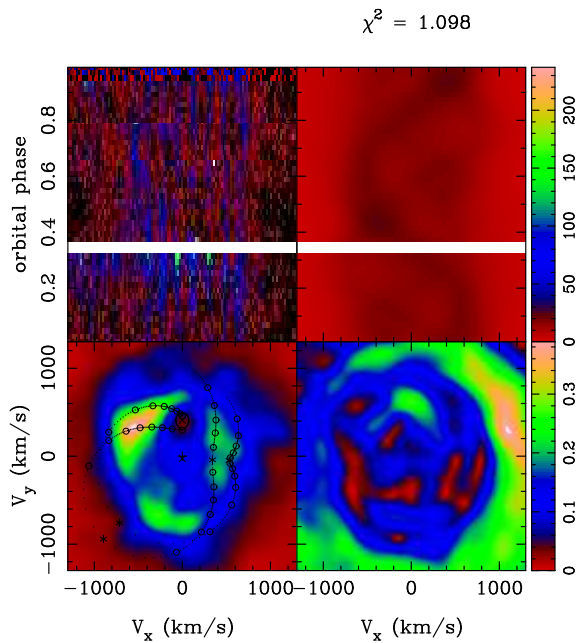}
\caption{The same as Figs.~\ref{fig:dopp1} and 7, but for Ca II (8662~\AA). The tomogram was created using an order of the MagE spectrum that was affected by sky-lines.}
\label{fig:dopp4}
\end{figure}

\section{Discussion}\label{section:disc}

\subsection{The hotspot of Nova Muscae}

Our tomograms display a bright spot which corresponds to the stream-disc interface. The presence of this feature runs counter to the hypothesis that Nova Muscae harbours an abnormally low rate of mass transfer compared to other black hole transients of comparable orbital period, and demonstrates the presence of a variable hotspot. 

By considering the position of the hotspot relative to the stream's trajectory, we identify the radial distance to the hotspot from the black hole. From the H~$\alpha$ Doppler map, the hotspot is found to lie squarely along the ballistic trajectory of the stream. The open circles along this line mark $10\%$ increments in $R_{\rm L1}$, which places the position of the hotspot near $R = 0.5 \pm 0.05R_{\rm L1}$. The H~$\beta$ Doppler map depicts the hotspot closer to the Keplerian trajectory, indicative of the disc's rotation along the course of the stream. This may represent post-shock H$~\beta$ emission as the stream is swept along into the disc. 

Assuming a Keplerian trajectory for the H~$\beta$ hotspot, we find that the emission originates from $R = 0.45 \pm 0.05R_{\rm L1}$, in good agreement with the H~$\alpha$ map. In the Ca II Doppler map, the hotspot appears sandwiched between the ballistic and Keplerian curves. The ballistic trajectory gives an upper limit for the hotspot at $R = 0.55 \pm 0.05R_{\rm L1}$ whereas the Keplerian lower limit corresponds to $R = 0.4 \pm 0.05R_{\rm L1}$. Despite the terminology, rather than envisioning a small ``spot'' of emission, one could instead envision the bright interface as a stretched patch reflecting a transition between ballistic and Keplerian trajectories, occurring at some distance $R \sim 0.5R_{\rm L1}$. In this case, the consistent distances suggest that the emission is relatively confined, but that different zones along this interface are responsible for the different line features. These zones, which are variously more or less pulled into the disc's motion, would naturally span the gamut from ballistic to Keplerian motion, while still originating from approximately the same location. For the location of such a hotspot, we find $R_{\rm hotspot} = 0.5 \pm 0.05R_{\rm L1}$.

In order to more readily place the location of the hotspot into context, we overlay the circularization radius of the system on the H~$\beta$ Doppler map. This radius corresponds to the disc orbit with the same specific angular momentum as the gas stream. Accordingly, the orbital velocity at the circularization radius is given by

\begin{equation}
V_{\rm circ} = \sqrt{\frac{GM}{R_{\rm circ}}},
\end{equation}

\noindent 
and radius $R_{\rm circ}$, given by \citet{Warner03} for $0.05<q<1$

\begin{equation}
R_{\rm circ} \approx 0.0859aq^{-0.426},
\end{equation}

\noindent 
where {\it a} is the binary separation.  For Nova Muscae, $R_{\rm circ} = (8.64 \pm 0.22) \times 10^{10}$ cm ($= 0.44 \pm 0.01R_{\rm L1}$), and its Keplerian velocity $V_{\rm circ}$ = 1001 $\pm$ 23 km s$^{-1}$. 
$V_{\rm circ}$ is marked on the H~$\beta$ Doppler map in Figure 9.  Although the tomogram indicates that the hotspot emission originates at smaller velocities than $V_{\rm circ}$ (i.e. exterior to $R_{\rm circ}$), it is noteworthy that it lies within 1~$\sigma$. Considering that the maximum distance to the disc edge is $\sim0.9R_{\rm L1}$ (\citealt*{Frank02}; eq. 5.122), Nova Muscae's hotspot, at $\sim0.5R_{\rm L1}$, is consistent with its circularization radius.

\begin{figure}
\begin{center}
\includegraphics[width=0.8\columnwidth]{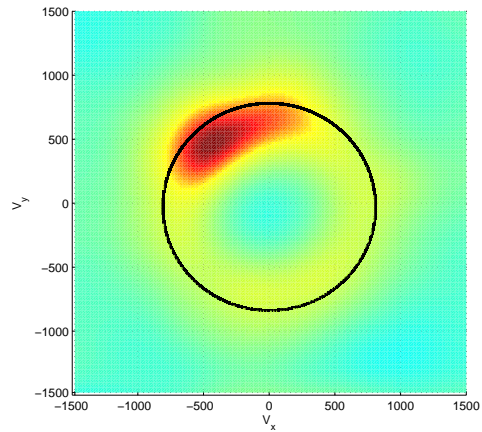}
\caption{The H~$\beta$ tomogram of Nova Muscae, shown close up. The black circle marks $V_{\rm circ}$sin($i$).}
\end{center}
\end{figure}

\subsection{Possible Stream-Disc Overflow}

SPH simulations performed by \citet{Armitage96} showed that a significant portion of the gas stream can be deflected from the outer edge of the disc and flow over the disc. The deflected material loses energy and settles at smaller radii approaching the circularization radius. A later analysis using SPH at higher resolution by \citet{Kunze01} revealed that a bow shock at the hotspot launches gas to about $20\degr - 25\degr$ above the disc plane.  This gas can produce absorption dips in the light curve at orbital phases 0.7--0.9 in systems with inclinations greater than 65$^{\circ}$--70$^{\circ}$. The simulations show the deflected mass settling back onto the disc by phase $\sim0.5$ independent of the mass ratio considered. In the right circumstances, such as during a spike in the mass transfer rate, or given a smaller sized disc, the simulations also show the overflow stream rebounding off the disc at phase 0.5 and producing yet another absorption dip in the light curve at phase 0.2. 


Our H~$\alpha$ (both IMACS in 2009 and MagE in 2010) and Ca II maps show a crescent shaped emission feature originating from the region stretching between phases $\sim 0.2-0.6$. The associated region is consistent with the point at which the deflected mass stream falls back onto the disc as in the \citeauthor{Kunze01} picture.  Together with the presence of the hotspot, this is suggestive of stream-disc overflow. The shift in velocity of the heated stream as evidenced by the hotspot drift from Figures 7-8 is consistent with the results of the \citeauthor{Kunze01} simulations as well.  

Within this overflow paradigm, one expects a diffuse cloud of absorbing gas to persist above the disc, reaching zenith at phase $\sim$0.7. Although the inclination of Nova Muscae, 54$^{\circ}$ \citep{Gelino01}, is likely too low to manifest corresponding absorption dips in its light curve, the absorbing matter is expected to be evident in H~$\alpha$ emission. Our H~$\alpha$ Doppler map also displays a gap in which the line emission is reduced around phase $\sim$0.7 and $\sim0.15$, the phases expected for absorption from mass overflow in the \citeauthor{Kunze01} picture. It is interesting to note that the H~$\alpha$ equivalent width variations shown by \citet{Casares97} in their figure 3 produce dips at phases $\sim$0.7 and $\sim$0.2\footnote{After accounting for a phase shift of 0.06 to bring their data up to date.}. Their and our observations are in good accord with the \citeauthor{Kunze01} simulations. 

In Figure 10 we present a rough schematic of the overflow process. We note that because velocity coordinates do not uniquely map into spatial coordinates, this figure is only approximate, based on one plausible realization of our tomograms.

\begin{figure}
\begin{center}
\includegraphics[width=1\columnwidth]{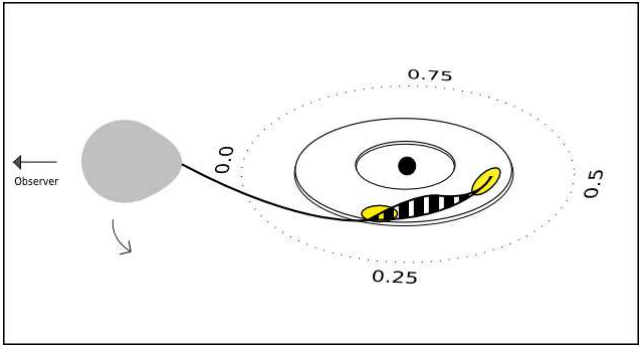}
\caption{A rough schematic of the stream-disc overflow in Nova Muscae. The numbers signify the phase of the system relative to the observer when the donor star reaches the marked position. The yellow spots indicate the hotspot where the accretion stream first hits the disc and the point of fallback of the deflected flow. The rise in elevation of the deflected flow above the disc is indicated by vertical lines.}
\end{center}
\end{figure}

\subsection{Detection of the Secondary Star}

While the Ca II emission from the disc is clearly present, our Ca II Doppler map also displays a compact region of lower emission coincident with the position of the secondary. We also observe fluctuation of the flux originating from a similar position in the modulation amplitude map. This is consistent with Ca II absorption from the secondary and is strongly reminiscent of the similar Ca II emission and absorption seen in XTE J1118+480 by \citet{Calvelo09}. However, further investigation is hindered by the presence of unsubtracted sky lines in our data. 

\citet{Casares97} detected emission in H~$\alpha$ from the region of the donor star, which they attributed to possible chromospheric activity. We detect a similar bright spot, but find that it is offset from the secondary at $\Delta V \sim$ 200~km s$^{-1}$ larger $V_y$ owing to the updated dynamical model. H~$\alpha$ emission, identified as originating from the secondary, has been observed in other black hole and neutron star systems (e.g., \citealt{Marsh94}; \citealt{Shahbaz99}; \citealt{Torres02}; \citealt{Gonz10}; \citealt{Peris12}), some compatible with the Roche lobe of the secondary and some slightly offset. Although it has been commonly attributed to chromospheric activity or X-ray heating of the secondary, an unequivocal interpretation has not yet been established. Our observation of similar H~$\alpha$ emission with a large offset is an intriguing result and we do not have an explanation for this as of yet.

\subsection{Comparison between Nova Muscae and A0620-00}

In this section we compare our tomograms with those of A0620-00 (\citealt{Neilsen08, Marsh94}), a system which is widely compared to Nova Muscae in the literature given the similarity of the system characteristics. The orbital period of Nova Muscae (P$_{\rm orb}$ = 10.38h) slightly exceeds that of A0620-00 (P$_{\rm orb}$ = 7.75h; \citealt{McClintock86}). When in outburst, A0620-00 has been suggested to harbour a slightly thicker and hotter disc compared to Nova Muscae \citep{Esin00}. 

Since both systems display prominent hotspots in Balmer emission, it is straightforward to compare them using the Doppler maps as a guide. To keep the comparison robust, we focus our comparison on examining the positions of the hotspots for both systems. We examine tomograms in H~$\alpha$, figure 10 in \cite{Neilsen08} and Figure 5 \& 6 here. As described in \S~4.1, we find a hotspot radius for Nova Muscae of $R = 0.5 \pm 0.05R_{\rm L1}$; for A0620-00, the H~$\alpha$ map reveals a larger but consistent radius, $R = 0.6 \pm 0.05R_{\rm L1}$. Assuming that the H~$\beta$ emission originates from circularized gas in Keplerian orbit, Nova Muscae's hotspot corresponds to a radius of $R = 0.45 \pm 0.05R_{\rm L1}$ and A0620-00's hotspot to $R = 0.5 \pm 0.05R_{\rm L1}$. These results are suggestive of a marginally smaller distance to Nova Muscae's hotspot, relative to the binary separation.

SPH simulations conducted by \citet{Foulkes04} showed that eccentric accretion discs manifest several characteristics in a Doppler tomogram. Notably, as the asymmetric disc rotates, the stream-disc impact point travels along a section of the edge of the disc creating an extended hotspot. Two distinct non-sinusoidal S-curves are then created by the disc gas and stream gas emission at the hotspot. They also predict crescent-shaped emission from the disc in the tomogram. An eccentric accretion disc should also show an offset of its centre of symmetry from the central position of the black hole. 

Although the H $\alpha$ tomogram in Nova Muscae displays a crescent shaped region of emission, we see no other compelling evidence for disc eccentricity. However, \citet{Neilsen08} finds compelling tomographic evidence that the disc in A0620-00 is eccentric. We suggest that unlike in the case of A0620-00 \citep{Neilsen08} the disc of Nova Muscae might not have grown large enough for the growth of eccentric modes. 

\section{Conclusions}

Our high resolution tomography of Nova Muscae shows a hotspot in both 2009 and 2010, confirming ongoing mass transfer via Roche-lobe overflow in the binary system during this epoch. Given its earlier absence, this raises the interesting possibility of a variable hotspot in Nova Muscae.

The Ca II maps display evidence for absorption from the secondary as well as emission from the disc. A similar observation for XTE J1118+480 \citep{Calvelo09} suggests that Ca II ($\lambda$8662) is of particularly high utility in constraining binary parameters in X-ray binary systems (see also \citealt{Spaandonk10}). Higher resolution spectra of the Ca II will help to better map the secondary in Nova Muscae.

Our tomograms support stream-disc overflow in the system. The bright spot in the lower half of the H~$\alpha$ map (also found in approximately the same location in the Ca II map) at phase $\sim$0.6 is interpreted as the point at which the deflected mass falls back to the disc. The overflow should result in an elevated cloud of absorbing gas above the disc, reaching its maximum altitudes at phase $\sim$0.7. Therefore, the presence of an absorption dip in the light curve at phase $\sim$0.7 would provide a ``smoking gun" for this interpretation. However, due to the moderate inclination of the system, the lack of a dip would not preclude this scenario. Such overflow might also explain the broader blue wing in the H~$\alpha$ profile observed by \citet{Orosz92} which was previously interpreted as an outflow of mass from the disc.

Bright spots in the phase region $\sim$0.5 - 0.75 have often been observed in other systems in Balmer and He I lines \citep{Marsh01}. We assert that these may be due to the ubiquitous presence of stream-disc overflow causing heating of the surface of the disc in the re-impact region. 

\section*{Acknowledgements}

We gratefuly acknowledge the use of the {\sc MOLLY} software package written by Tom Marsh. We would like to thank Jorge Casares for sharing his data on Nova Muscae with us. We would also like to thank the referee for helpful comments. JFS has been supported by NASA Hubble Fellowship grant HST-HF-51315.01. LCH acknowledges support from the Kavli Foundation, Peking University, and the Chinese Academy of Science through grant No. XDB09030102 (Emergence of Cosmological Structures) from the Strategic Priority Research Program. 

\bibliography{biblio}

\section*{Appendix}

Figure 11 shows a finding chart of Nova Muscae created using a 60~s image obtained using the Bessel V filter on IMACS ($\sim$0.85 arcsec seeing).

\begin{figure*}
\begin{center}
\includegraphics[width=110mm]{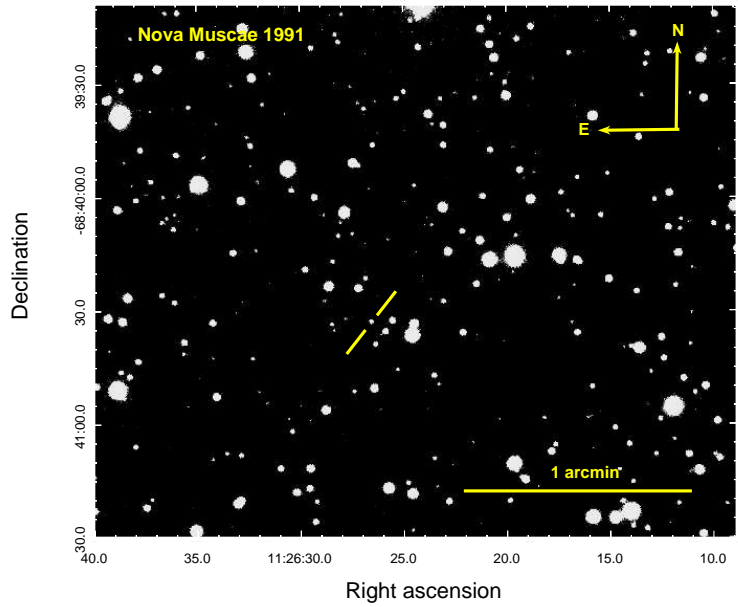}
\caption{The field of Nova Muscae 1991 with the system marked by two lines.}
\end{center}
\end{figure*}

\label{lastpage}

\end{document}